\documentclass[runningheads]{llncs}
\usepackage[T1]{fontenc}
\usepackage{graphicx}
\usepackage{xcolor}
\usepackage[dvipsnames]{xcolor}

\usepackage{eso-pic}
\usepackage{graphicx}

\newcommand{\addicprtopnote}{%
  \AddToShipoutPictureBG{%
    \AtPageUpperLeft{%
      \raisebox{-60pt}{
        \parbox{\paperwidth}{\centering\Large Submitted to ICPR 2026}%
      }%
    }%
  }%
}

\addicprtopnote

\begin{document}
\title{Parameter-Efficient Deep Learning for Ultrasound-Based Human-Machine Interfaces \thanks{Supported by IMEC.}}
%
%
\author{Antonios Lykourinas\inst{1} \and
Chinmay Pendse\inst{2,3}\and
Francky Catthoor \inst{4} \and 
Veronique Rochus\inst{3} \and Xavier Rottenberg \inst{3} \and Athanassios Skodras\inst{1}}
\authorrunning{A. Lykourinas et al.}
\titlerunning{Parameter-Efficient Deep Learning for Ultrasound-Based HMIs}

%
\institute{Dept. of ECE, University of Patras, Patras, Greece \\
\email{\{alykourinas,skodras\}@\{ac.upatras,upatras\}.gr} \and 
Katholieke Universiteit (KU) Leuven, Leuven, Belgium  \and
Imec, Leuven, Belgium \\ 
\email{\{Chinmay.Pendse,Veronique.Rochus,Xavier.Rottenberg\}@imec.be}\and
Dept. of ECE, National Technological University of Athens, Athens, Greece \\
\email{catthoor@microlab.ntua.gr}
}

\maketitle              
\begin{abstract}

Ultrasound (US) has emerged as a promising modality for Human-Machine Interfaces (HMIs), with recent research 
efforts exploring its potential for Hand Pose Estimation (HPE). A reliable solution to this problem could
introduce interfaces with simultaneous support for up to 23 degrees of freedom encompassing all hand and wrist 
kinematics, thereby allowing far richer and more intuitive interaction strategies. Despite these promising 
results, a systematic comparison of models, input modalities and training strategies is missing from the 
literature. Moreover, there is only one publicly available dataset, namely the Ultrasound Adaptive Prosthetic 
Control (Ultra-Pro) dataset, enabling reproducible benchmarking and iterative model development. In this paper, 
we compare the performance of six different deep learning models, selected based on diverse criteria, on this 
benchmark. We demonstrate that, by using a step learning rate scheduler and the envelope of the RF signals 
as input modality, our 4-layer deep UDACNN surpasses XceptionTime's performance by $2.28$ percentage 
points while featuring $87.52\%$ fewer parameters. This result ($77.72\%$) constitutes an absolute 
improvement of $0.88\%$ from previously reported baselines. According to our findings, the 
appropriate combination of model, preprocessing and training algorithm is crucial for 
optimizing HMI performance. 

\end{abstract}
\section{Introduction}

In daily life, humans interact with machines through interfaces such as keyboards, mice, 
touch screens, buttons, and controllers. We refer to them collectively by using the umbrella 
term Human-Machine Interfaces (HMIs). Beyond their role in everyday human-machine interaction, 
HMIs have been introduced into a wide range of domains, including rehabilitative and assistive 
technologies in medicine, robotic arm control and tele-operation in industry, and extended reality 
controllers in entertainment \cite{Esposito2021}. Nevertheless, most conventional HMIs suffer 
from communication bandwidth limitations, as their operation is restricted to limited 
Degrees-of-Freedom (DoFs). For example, a mouse cursor operates in only 2 DoFs. In contrast, 
our hands are capable of performing dexterous movements, spanning 23 DoFs (20 hand $+$ 3 wrist DoFs), 
enabling far richer and natural interaction strategies. 

Hand Pose Estimation (HPE), i.e., the continuous estimation of hand joint angles, has been approached 
using several sensing modalities, including optical sensors (RGB, depth, Motion Capture), Inertial Measurement 
Unit (IMU) sensors, and wearable gloves as a means of generating robust control signals for HMIs. However, 
optical-based approaches often require cumbersome multi-view camera setups for effective inference, 
and are sensitive to occlusion and lighting conditions, while wearable gloves can restrict natural 
movements \cite{salter2024}. In contrast, non-invasive biosignal modalities, such as surface-electromyography 
(sEMG) and ultrasound (US), not only overcome the aforementioned limitations but also provide the only 
viable means of generating control signals for amputees, who number approximately $2.3$ million adults 
in the USA alone \cite{Rivera2024}.

Within the biosignal HMI community, the Hand Gesture Recognition (HGR) problem is commonly simplified to 
Static Hand Pose Recognition (SHPR), as this formulation covers several practical applications. 
However, as evidenced by sEMG research, even SHPR remains a challenging problem that has required years 
of extensive iterative model development to substantially improve performance. As a reference point, the 
first Deep Learning (DL) baseline released with the NinaPro database in 2014 reached an accuracy of only 
$60.70\%$ \cite{Atzori2014}. Recent approaches, such as TraHGR \cite{zabihi2022}, have reported an accuracy 
of $84.13\%$ by utilizing a $100$ ms temporal window. Building on prior research, recent industry efforts 
have introduced large-scale datasets targeting complex tasks such as keyboard typing \cite{sivakumar2025} 
and HPE \cite{salter2024}, attempting to push generalization boundaries across sessions, subjects, and 
gestures. 

In recent years, US has emerged as a promising modality for HMIs due to its ability to provide 
musculoskeletal structure information with sub-millimeter precision and high temporal resolution 
\cite{YangReview2024}. US has been extensively used to tackle the SHPR problem, reporting accuracies 
surpassing sEMG literature results, under similar experimental setups \cite{Lian2024,Vostrikov2024,Yang2018,Zeng2023}. 
Furthermore, as demonstrated in several hybrid HMI studies \cite{spacone2025,Wei2023}, US complements 
the electrophysiological nature of sEMG signals, offering performance enhancements when fused. In addition, 
the superiority of US becomes more evident in simultaneous and proportional control (SPC) tasks where US, 
compared to sEMG, has been able to maintain high coefficient of determination $R^{2}$ values for 3 DoFs 
even during their simultaneous co-activation \cite{Sgambato2023}. Finally, a recent study has explored 
the potential of US for tackling the challenging problem of HPE \cite{spacone2025}.

Despite these promising results, the broader US-based HMI community faces challenges that limit the 
field's overall progress. To begin with, most studies are conducted on private datasets, often serving 
as a proof-of-concept for novel US acquisition systems \cite{Frey2022,Sgambato2024,Sgambato2023,Xia2019}. 
In addition, the existence of only a single publicly available dataset, i.e. the Ultrasound Adaptive 
Prosthetic Control (Ultra-Pro) dataset, renders direct one-to-one comparison with other works infeasible 
due to slightly different problem formulation, experimental setups, and evaluation strategies. As a 
consequence, opportunities for iterative model development and benchmarking across methods are limited. 
Finally, an often overlooked aspect is the exploration of alternative modalities, as most studies adopt 
standard preprocessing procedures followed in medical US imaging, with a few notable exceptions \cite{Fernandes2021,Gao2024}. 

Inspired by sEMG research advancements, we aim to improve the low baselines currently reported on 
the Ultra-Pro dataset. This experimentation is of utmost importance before scaling to large-scale 
datasets targeting complex tasks (e.g. HPE, keyboard typing), where extensive exploration will become 
much more resource- and time-intensive. In this work, we compare six deep learning (DL) models, 
covering diverse architectural design choices (from attention-based to convolution-based) drawn 
both from state-of-the-art US-based HMI and time-series classification literature, on the Ultra-Pro 
dataset. Our goal is to identify optimal models in terms of performance and parameter efficiency 
for US-based HMIs and to assess whether the primary limiting factor of performance is input modality, 
model architecture, or the training algorithm. Moreover, despite its small scale, Ultra-Pro provides 
a challenging benchmark, since it introduces an additional DoF, i.e. wrist rotation, into the SHPR 
problem. This SHPR extension enables improved prosthesis functionality for transradial amputees, 
since the combination of the selected set of static hand poses with wrist rotation reproduces hand 
kinematics commonly performed in Activities of Daily Living (ADL) \cite{Yin2025}. Our experimentation 
lead us to a key finding: the performance of US-based HMIs is jointly affected by multiple factors 
such as the choice of the input modality, network, and training algorithm. Our main contributions 
can be summarized as follows:

\begin{enumerate}
    \item We conduct extensive Hyperparameter Optimization (HPO) studies on six DL models,
    covering both state-of-the-art US-based HMI and time-series classification literature. 
    Additionally, we provide analysis of the studies, discuss hyperparameter importance and 
    finally, derive optimal configurations for the task of SHPR with concurrent wrist rotation. 
    \item We show that for standard preprocessing, XceptionTime \cite{rahimian2019} surpasses 
    the performance of all other models  on the Ultra-Pro intra-session benchmark, 
    reaching an accuracy of $75.44\%$. 
    \item We show that the correct combination of model architecture, input modality and training 
    algorithm is crucial for optimizing performance. In particular, XceptionTime \cite{rahimian2019} 
    (selected for its strong performance) and UDACNN \cite{Lykourinas2024} (selected for its 
    parameter efficiency) responded differently to input modality and training algorithm 
    modifications. By using a step learning rate scheduler and the envelope of the RF signals 
    as the input modality, UDACNN outperformed XceptionTime by $2.28$ percentage points while 
    featuring $87.52\%$ fewer parameters, reaching a $77.72\%$ recognition accuracy on the 
    Ultra-Pro intra-session benchmark. This result constitutes an absolute improvement
    of $0.88\%$ from the original Ultra-Pro baselines, obtained under a well-defined reproducible
    benchmark, addressing several missing implementation details from the original study.
\end{enumerate}

The rest of the paper is organized as follows:  Section 2 introduces the methods used 
in our experiments. Section 3 provides a brief description of the Ultra-Pro dataset 
and our experimental setup. In Section 4, we present experimental results and finally, 
Section 5 concludes our work.

\section{Methods}

\subsection{Models}

In this paper, we evaluate the performance of six DL models on the Ultra-Pro dataset. Model 
selection was based on multiple criteria. First and foremost, we included models that have
been previously benchmarked on Ultra-Pro \cite{Lykourinas2024,Yang2022}. Furthermore, we included 
models that have demonstrated strong performance in US-based SHPR, under experimental setups 
similar to Ultra-Pro \cite{Lian2024,Zeng2023}. In addition, we included the XceptionTime architecture 
due to its exceptional performance in sEMG-based recognition \cite{rahimian2019} and biomedical 
time-series applications \cite{Doldi2022,Singh2023}. Finally, we included a Vision Transformer 
(ViT)\cite{dosovitskiy2021} because it represents an entirely different learning paradigm compared 
to the previous convolutional-based approaches.

\begin{enumerate}
    \item A ViT variant \cite{dosovitskiy2021} tuned for US-Based HGR. In our implementation, we
    use fixed 2D sinusoidal positional embeddings and average the final patch representations of 
    the transformer encoder for classification, which is performed by a single-layer linear head.
    \item The A-mode Ultrasound Network (AUSNet), proposed as part of an adaptive learning algorithm 
    to mitigate performance downgrades caused by muscle fatigue \cite{Zeng2023}.
    \item The Multi-branch Squeeze-and-Excitation Network (Multi-branch SE Net) proposed in a transfer learning 
    study \cite{Lian2024}, where A-mode US signals are processed in parallel branches and features are
    concatenated channel-wise after the SE mechanism is applied. A two-layer MLP network is used for 
    the final gesture prediction.
    \item The XceptionTime model, originally proposed for sEMG-based HGR \cite{rahimian2019}.
    \item The Single-Task Convolutional Neural Network (STCNN), introduced in the original Ultra-Pro study \cite{Yang2022}.
    \item The CNN model (UDACNN) proposed in our previous inter-session re-calibration work \cite{Lykourinas2024}.
\end{enumerate}

For additional details, readers could refer to the original publications. 

\subsection{Tree-Structured Parzen Estimators}

In our experiments, we used Tree-structured Parzen Estimators (TPE) for HPO. In contrast to Grid and Random 
Search, which do not incorporate feedback during optimization, TPE exploits information from previous evaluations 
to guide the search towards promising regions of the search space. This property makes TPE effective in large 
search spaces. This behavior was also observed during our ViT experimentation, where TPE outperformed Random 
Search by a significant margin of $3$ percentage points under an identical budget of 500 evaluations. As a 
result, TPE was adopted in all subsequent HPO studies using settings recommended in a prior work \cite{watanabe2023tree}.

TPE is a Bayesian optimization algorithm designed to efficiently explore complex, potentially tree-structured 
search spaces, i.e. search spaces with conditional parameters, and has been shown to be effective in practical 
HPO tasks \cite{watanabe2023b}. The algorithm consists of two main phases: a warm-up phase and the main optimization 
loop. In the warm-up phase, hyperparameter configurations $\mathbf{x}$ are randomly sampled from the search space 
and evaluated to collect initial observations $D=\{(\mathbf{x}_{i},y_{i})\}_{i=1}^{N_{init}}$, where $y_{i}$ is 
the objective value for a given configuration $\mathbf{x}_{i}$. In the main loop, observations are split into two 
groups: good configurations $D^{(l)}$ (top-performing fraction) and bad configurations $D^{(g)}$ (rest) and parzen 
estimators are used to model the distributions $p(\mathbf{x}|D^{(l)})$ and $p(\mathbf{x}|D^{(g)})$. The two 
distributions are used to construct the \textit{acquisition function}, defined as their ratio, from which we sample 
candidates that are likely to improve the objective. The best candidate is included in our observations $D$ and 
this process is repeated until the pre-defined budget is exhausted.

\subsection{Learning Rate Schedulers}

In this work, we also explored the use of learning rate schedulers for improving the convergence 
of our training algorithms. Learning rate schedulers have become an important tool in DL, since 
they combine both the benefits of small and large learning rates by dynamically adjusting the learning 
rate throughout the training process. We restricted our experimentation to two representative schedulers: 
an \textit{exponential scheduler}, which decays the learning rate by a factor $\gamma$ every epoch, 
and a \textit{step scheduler}, which reduces the learning rate by a factor $\gamma$ every $s$ epochs. 
Our scheduler choices are motivated by two reasons: \textit{i)} they involve few hyperparameters, 
making them easier to debug and tune and \textit{ii)} despite their simplicity, they have been proven 
very effective in various domains \cite{rahimian2019}. 

\section{Experimental Setup}

\subsection{Ultra-Pro}

The Ultra-Pro dataset contains US, sEMG, and IMU recordings from four transradial amputee 
subjects \cite{Yang2022}. As mentioned in the Introduction, Ultra-Pro 
is the only publicly available dataset and targets the task of simultaneous finger gesture recognition and
wrist rotation angle estimation, which constitutes a more challenging benchmark than standard 
SHPR formulations. Furthermore, its public availability, which enables reproducible and systematic 
comparisons across methods, combined with its practical relevance for functional prosthesis 
applications, renders Ultra-Pro the \textit{de facto} benchmark dataset within the US-based 
HMI community. The dataset includes six distinct finger gestures, namely Rest (RS), 
Power Grip (PG), Fine Pinch (FP), Index Point (IP), Tripod Grip (TG), and Key Grip (KG), selected 
based on ADL and prosthetics literature. Each subject participated in three distinct sessions. In 
each session, subjects executed the gestures sequentially. Gesture execution involved maintaining 
a static hand pose for 50 seconds while concurrently performing wrist rotations at an approximate 
frequency of $0.5$ Hz. Ultrasound data were collected using a customized armband equipped with $8$ 
evenly spaced transducers around the forearm operating at $5$ MHz, driven by a dedicated acquisition 
device \cite{Yang2021}. Data from all modalities were captured simultaneously and synchronized 
through specialized software. For more details, we encourage readers to refer to the original 
publications \cite{Yang2021,Yang2022}.

\subsection{Data Preprocessing and Preparation}

In this paper, we adopt two preprocessing configurations in order to derive two distinct input modalities 
for our networks. For our primary configuration, we follow the standard procedure of Time-Gain Compensation 
(TGC), bandpass filtering, envelope detection and log compression (TGC and log compression parameters were 
taken from \cite{Yang2021Decoding}). Additionally, only in our final experiment, we investigate an alternative 
modality in which we directly extract the envelope of the raw RF signals. We refer to these preprocessing 
configurations as A-mode US and Envelope(RF). In both cases, after discarding the first and last two samples, 
the pre-processed signals from all channels ($C$) are vertically concatenated to form the network input of 
size $C\times L$, where $L$ is the sequence length. Finally, for fair comparison, we followed the same 
train-validation-test split as in the original paper \cite{Yang2022}. 

\subsection{Training Details}

The task of SHPR is formulated as a classification problem. Thus, we use cross entropy loss 
as our loss function. Model parameters are optimized using the Adam optimizer with default 
settings ($\beta_{1}=0.9,\beta_{2}=0.999$). Adam was selected because \textit{i)} it was 
already used in the original papers and, \textit{ii)} it is a reliable starting point for 
gradient-based optimization of deep learning architectures when no optimizer is specified. 
Finally, we chose to use a batch size of 32 throughout our experiments to balance training 
time with the beneficial stochasticity of gradient updates, which often improves convergence.

\section{Results} 

This section outlines the structure of our experiments and the motivation behind them. Our study
was conducted in three stages:

\begin{enumerate}
    \item HPO studies for the six DL models.
    \item Intra-session model performance assessment, i.e. training and testing are carried out independently 
    for each session, followed by identification of two candidates based on performance and parameter efficiency 
    for further experimentation.
    \item Exploration of the effect of learning rate schedulers and input modality on the selected 
    candidates.
\end{enumerate}

Except for the ViT experimentation, where we had to identify optimal hyperparameters from scratch, the motivation 
behind the HPO studies is two-fold: \textit{i)} identify whether the hyperparameters reported in the original papers 
are optimal for the Ultra-Pro experimental setup and \textit{ii)} infer missing hyperparameters and training details 
that were not specified in the original papers. The HPO studies ensured a fair comparison on the Ultra-Pro intra-session benchmark, where we identify the two promising candidates (based on performance and parameter efficiency) 
for further experimentation. In our final experimental section, we investigate whether performance limitations arise 
from the training algorithm and the input modality itself, rather than the underlying model architecture. Finally, 
it is important to note that each individual experiment (except HPO studies) was conducted 10 times using different
random seeds.

\subsection{HPO Results}

This section presents the HPO results for all six DL models. For consistency, all HPO studies were conducted 
using TPE with a budget of 500 trials and 50 warm-up trials. Each experiment follows the same structure: 
\textit{i)} definition of the hyperparameter search space, \textit{ii)} analysis of promising configurations 
and hyperparameter importance discussion, and \textit{iii)} selection of the final configuration based on 
step \textit{ii}. 

\subsubsection{HPO results for the ViT variants}

The search space required careful design thus we restricted most hyperparameters to categorical variables, 
except for the learning rate and the number of epochs, for two reasons: \textit{i)} only certain configurations 
can produce valid architectures (e.g. Patch Embedding (PE) dimension/image width must satisfy divisibility constraints 
with respect to the number of heads/patch width) and \textit{ii)} discretizing the space reduces its effective 
cardinality, enabling a more efficient search. The final search space can be seen in Table \ref{tbl:USViT_search_space}. 
Note that \textit{FFN mul} refers to the feed-forward expansion ratio,  i.e., the multiplier applied to the hidden 
dimension of the MLP layers.

\begin{table}
    \centering
    \caption{ViT: Hyperparameter Search Space for TPE HPO study}
    \label{tbl:USViT_search_space}
    \begin{tabular}{|l|l|l|}
         \hline 
         Hyperparameter & Variable Type & Range \\
         \hline  
         Learning rate & float & $[1e-5,1e-1]$  \\
         PE dimension & categorical & $[32,48,64,128,256,512,1024]$ \\
         Patch height & categorical & $[1,2,4,8]$ \\ 
         Patch width & categorical & $[4,8,16,32,64,120,240,480]$  \\
         Heads & categorical & $[2,4,8,16]$ \\
         Encoder blocks & categorical & $[1,2,3,4,5,6]$ \\
         FFN mul & categorical & $[1,2,3,4]$  \\
         Epochs & Integer & $[1,200]$ \\ 
         \hline
    \end{tabular}
\end{table}

\begin{figure}
    \centering
    \includegraphics[width=\textwidth]{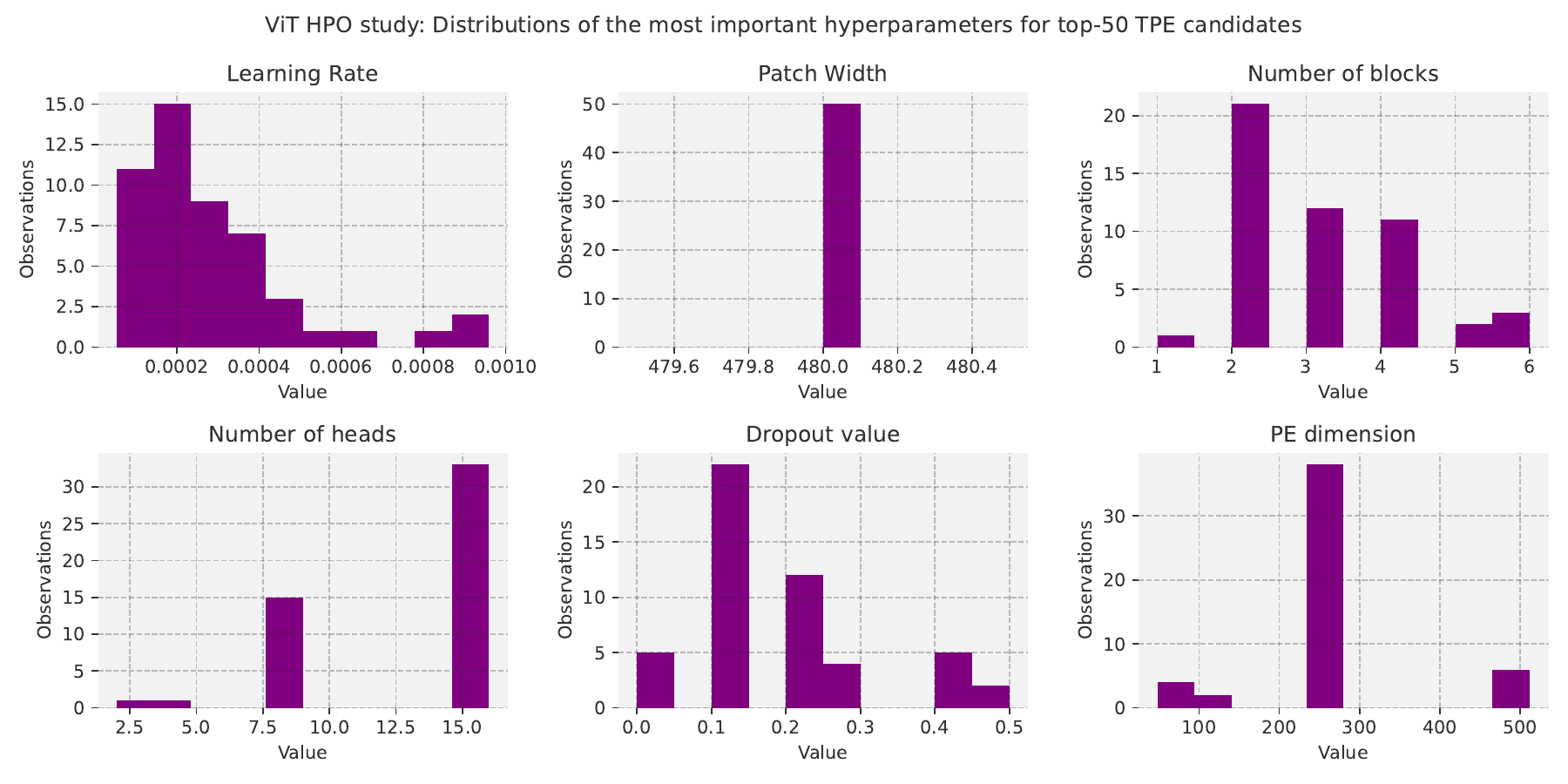}
    \caption{ViT HPO Study: Distributions of the most important hyperparameters based on the top-$50$ performing
    candidates sampled by TPE}
    \label{fig:ViT_Variants_HP_Distributions}
\end{figure}

\begin{figure}
    \centering
    \includegraphics[width=0.5\linewidth]{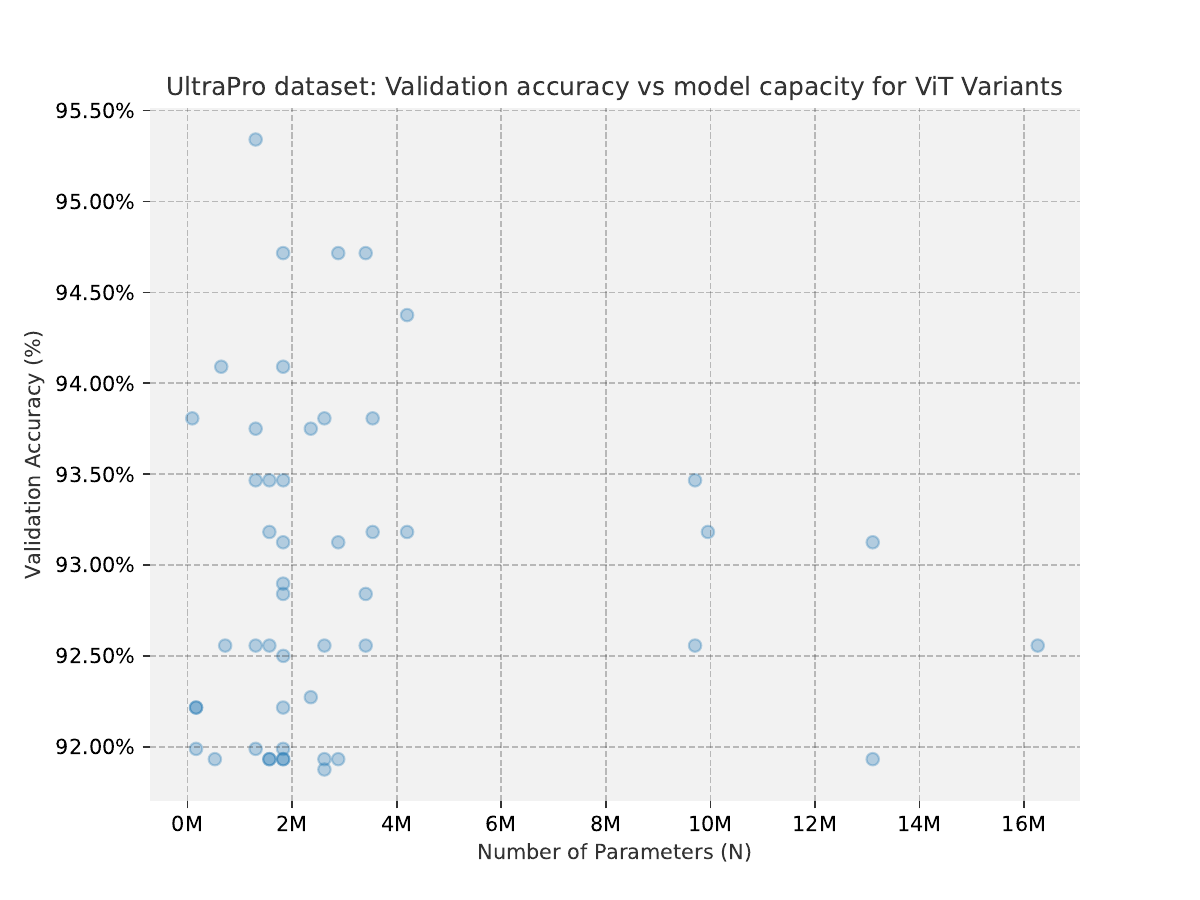}
    \caption{ViT HPO Study: Scatter plot between the number of parameters and HPO objective (Validation Accuracy) for the
    top-$50$ performing candidates sampled by TPE.}
    \label{fig:ViT_variants_val_acc_vs_n_parameters}
\end{figure}

Based on our HPO analysis, we came up with the following observations:

\begin{enumerate}
    \item Patch width emerged as the most influential hyperparameter as TPE's top-50 candidates were 
    solely sampled with the largest patch width available in the search space (See Figure \ref{fig:ViT_Variants_HP_Distributions}),
   \item TPE also showed a clear preference towards ViTs with a relatively shallow architecture, typically 
   between 2 and 4 encoder blocks.
   \item Among the top-50 performing candidates, TPE predominantly sampled a PE dimension of 
   256, 16 attention heads, and a dropout rate of $0.1$ (See Figure \ref{fig:ViT_Variants_HP_Distributions}). 
   \item Over-parameterized ViT variants underperformed, indicating that increased model capacity
   does not necessarily improve performance. Best performing variants typically fell within a parameter 
   range of approximately $200k$ to $3M$ (See Figure \ref{fig:ViT_variants_val_acc_vs_n_parameters})
\end{enumerate}

The selected configuration derived from our HPO study is shown in Table \ref{tbl:optimal_ViT_Config}. This model
has $647,814$ trainable parameters and will be referred to as USViT throughout the paper. In subsequent experiments,
USViT was trained using a learning rate of $0.00048$ for $57$ epochs.

\begin{table}
    \caption{Optimal ViT Configuration discovered through our HPO study analysis}
    \label{tbl:optimal_ViT_Config}
    \centering
    \begin{tabular}{|l|l|l|l|l|l|}
       \hline
       Parameter  & Patch Size & PE dimension & Heads & Encoder blocks & Dropout \\
       \hline
       Value      & $(2,480)$ & $256$ & 16 & 3 & 0.1  \\
       \hline 
    \end{tabular}
\end{table}

\subsubsection{HPO results for AUSNet} 

The hyperparameter search focused on dropout, learning rate and number of epochs for 
training (See Table \ref{tbl:AUSNet_search_space}). Hyperparameters missing from the 
original paper \cite{Zeng2023}, e.g. learning rate and number of epochs, were inferred 
through TPE Optimization. 

\begin{table}
    \centering
    \caption{Search Space for the AUSNet Network}
    \label{tbl:AUSNet_search_space}
    \begin{tabular}{|l|l|l|}
         \hline
         Hyperparameter & Type & Range  \\
         \hline 
         Learning Rate & Float & [$1e-5$, $1e-1$] \\
         Dropout & Categorical & $[0.0,0.1,0.2,0.3,0.3,0.5]$\\
         Number of Epochs & Integer & $[1,200]$ \\
         \hline
    \end{tabular}
\end{table}

Based on our analysis, we came up with the following observations:

\begin{enumerate}
    \item Convergence was achieved across a wide range of learning rate values, with the best-performing 
    candidates consistently falling within the range between $0.005$ to $0.01$.
    \item Both dropout values of $0.2$ and $0.5$ supported good generalization. 
\end{enumerate}

The hyperparameter configuration derived from the HPO study is a learning rate of $0.0064$, a dropout rate
of $0.2$ and $99$ training epochs.

\subsubsection{Remaining HPO Studies Results} We briefly summarize the results from the remaining HPO studies:

\begin{itemize}
    \item \textbf{Multi-branch SE Net:} The search focused on learning rate, dropout, number of epochs, and SE 
    reduction ratio. Unable to discover hyperparameter configurations with satisfactory performance (best trial $86.65\%$). 
    This could be attributed to the transducer operating frequency mismatch ($2.25$ MHz vs. $5$ MHz used in all 
    other studies), suggesting that model parameter adaptation to the new experimental setup is required. 
    \item \textbf{XceptionTime:} The search space included solely learning rate and number of epochs. 
    Hyperparameter configuration discovered with TPE: a learning rate of $0.0014$ and $60$ training epochs.
\end{itemize}

The learning rates for STCNN and UDACNN were adopted from \cite{Lykourinas2024}. For both models, a batch size of
$32$ was used. Finally, STCNN and UDACNN were trained for $30$ and $60$ epochs, respectively.

\subsection{Intra-session Results}

In this section, we assess the intra-session gesture recognition performance of all succesfully tuned models 
by means of Classification Accuracy (CA) on the Ultra-Pro dataset. In contrast to the original study \cite{Yang2022},
we use our own intra-session benchmark, which is well-defined and simulates a realistic use case where the HMI 
needs to be operable shortly after user data have been collected \cite{Lykourinas2024}. Table \ref{tbl:Ultra_Pro_avg_ca} 
summarizes the overall results.

\begin{table}
    \centering
    \caption{Ultra-Pro Dataset: intra-session benchmark results. Models are presented in descending
    order based on their average CA.}
    \label{tbl:Ultra_Pro_avg_ca}
    \begin{tabular}{|l|l|l|l|l|l|}
        \hline
        Model & XceptionTime & USViT & UDACNN & AUSNet & STCNN  \\
        \hline
        Average CA & $75.44\%$ & $74.54\%$ & $74.03\%$ & $71.27\%$ & $71.26\%$ \\
        Trainable parameters & 405,250 & 647,814 & 50,584 & 733,572 & 418,806 \\
        \hline 
    \end{tabular}
\end{table}

XceptionTime emerged as the top-performing model, achieving a accuracy of $75.44\%$ across all
subjects and sessions. Its strong performance is due to robust inter-subject generalization,
particularly evident in the sessions of the third subject, which were identified as the most 
challenging. We attribute this ability to the use of multiple parallel depthwise convolutions 
in each XceptionTime module, which capture temporal patterns at multiple receptive fields 
without requiring pre-defining kernel sizes. 

UDACNN achieves competitive performance, trailing behind XceptionTime and USViT by $1.41$ and $0.51$ 
percentage points, respectively, while using $87.52\%$ and $92.19\%$ fewer trainable parameters. 
Surprisingly, STCNN and AUSNet achieve significantly lower performance than UDACNN despite having 
$14.50\times$ and $8.28\times$ the number of its trainable parameters, respectively. This demonstrates 
that careful model design is more important than simply increasing capacity which can lead to overfitting. 

\subsection{Experimentation with Schedulers and Input Modality}

In this section, we experiment with schedulers and input modality in an attempt to further improve
the performance of XceptionTime and UDACNN models. XceptionTime was selected due to its strong and 
consistent inter-subject performance, whereas UDACNN was chosen for its high parameter efficiency. 
We first examine the impact of learning rate schedulers on the convergence behavior of both models. 
Additionally, we evaluate Envelope(RF) as an alternative input modality to assess whether the standard 
A-mode US modality constitutes a limiting factor in model performance. 

\subsubsection{Experimentation Phase for UDACNN} 

By manually tuning the learning curves across different subjects and sessions of the Ultra-Pro dataset, we
identified the optimal scheduler configurations. For the exponential scheduler, we used a decay factor 
$\gamma=0.9$ whereas for the step scheduler, we used a step size $s=10$ and a decay factor $\gamma=0.5$. These 
scheduler configurations were adopted for both modalities. For both schedulers, an initial learning rate of 
$0.003$ and $60$ epochs were used for training. 

\subsubsection{UDACNN Performance Comparison} 

We evaluate the impact of learning rate schedulers and input modality on the performance of UDACNN. Specifically, 
we compare training with no scheduler, an exponential scheduler, and a step scheduler, using both A-mode US and 
Envelope(RF) modalities. The average classification accuracies (followed the intra-session evaluation procedure) 
on the Ultra-Pro dataset are reported in Table \ref{tbl:scheduler_results}.

\begin{table}
    \caption{UDACNN: Average CA on the Ultra-Pro intra-session benchmark for different 
    experimental configurations}
    \label{tbl:scheduler_results}
    \centering
    \begin{tabular}{|l|l|l|l|}
        \hline
                        &  None       & Exp. Scheduler  & Step Scheduler \\
        \hline
         A-line Signals &  $74.03\%$   & $75.29\%$      &   $76.51\%$    \\
         \hline
         Envelope(RF)   &   $75.33\%$  & $77.32\%$      &   $77.72\%$     \\
         \hline
    \end{tabular}
\end{table}

Based on these results, we came up with the following observations:

\begin{enumerate}
    \item UDACNN benefits from the use of a learning rate scheduler, regardless of the input modality.
    \item The Envelope(RF) modality leads to improved generalization compared to A-mode US, suggesting that 
    additional pre-processing steps omitted in the standard configuration might be redundant or even have a 
    negative effect on the performance of this model. 
    \item Combining the step scheduler with Envelope(RF) modality achieves an average recognition accuracy of 
    $77.72\%$, surpassing results reported in the original work by approximately $1\%$, in which several
    implementation details are missing \cite{Yang2022}.
\end{enumerate}

\subsubsection{UDACNN: Inter-subject compatibility of Exponential Scheduler Parameters} To assess whether the
performance gains of UDACNN depend on subject-specific tuning, we evaluated the inter-subject robustness of the 
exponential learning rate scheduler. Session-wise hyperparameter optimization studies revealed consistent preferences 
for a decay factor above $0.75$ across all subjects, indicating limited sensitivity to inter-subject variability. 
The search space also included $\gamma=1.0$ (i.e., no decay) to allow TPE identify whether learning rate 
scheduling was beneficial. The manually selected hyperparameters align well with the optimal configurations 
observed across sessions supporting their use as a robust, subject-agnostic choice. 

\subsubsection{XceptionTime: Effect of Schedulers \& Pre-processing} We evaluated the impact of learning rate 
schedulers and Envelope(RF) input modality on XceptionTime. Neither exponential nor step scheduler improved 
performance using A-mode US as input. Furthermore, when we used Envelope(RF), the average recognition accuracy 
on Ultra-Pro decreased from $75.44\%$ to $74.77\%$. The degradation in performance when using the Envelope(RF) 
modality is potentially attributed to the loss of relative signal contrast across the dynamic range. Logarithmic 
compression amplifies lower-amplitude sections of the signal while compressing higher-amplitude ones. This
process emphasizes subtle variations in the envelope signal that might be crucial for the XceptionTime's 
convolutional filters to extract discriminative features. In contrast, Envelope(RF) preserves the full dynamic 
range, which makes smaller but more informative patterns less distinguishable, leading to poorer model performance. 
Furthermore, this observation is consistent with the original XceptionTime study \cite{rahimian2019}, where the 
application of the $\mu$-law transform (another non-linear transform used for dynamic range compression) to 
the sEMG signals led to a substantial performance improvement, even increasing by $10\%$ when temporal windows 
of $50$ ms were used.

\section{Conclusions}

In this paper, we conducted a comprehensive study of six DL models, covering standard time-series 
classification approaches as well as models explicitly designed for US-based HGR, and assessed their 
performance on the publicly available Ultra-Pro dataset. We performed extensive HPO using TPE to derive 
optimal configurations for each architecture and analyzed the influence of crucial hyperparameters on 
model performance. Furthermore, we leveraged those findings to evaluate the task-refined models in the 
intra-session benchmark of Ultra-Pro, where training and testing are carried out independently 
for each session. Finally, we selected two promising models, XceptionTime, for its strong performance 
and UDACNN, for its parameter efficiency and explored strategies to further improve their performance 
by \textit{i)} incorporating learning rate scheduling and \textit{ii)} examining the envelope of the RF 
signals as an alternative input modality. Our results indicated that when trained with vanilla Adam 
optimizer, XceptionTime provides the strongest baseline, achieving an overall accuracy of $75.44\%$, 
followed by USViT and UDACNN with accuracies of $74.54\%$ and $74.03\%$, respectively. However, by 
applying a step scheduler and using the envelope RF as input we substantially improved UDACNN's 
performance, raising its accuracy to $77.72\%$. This result constitutes an absolute improvement of 
$0.88\%$ over the original Ultra-Pro baseline (for single-task models), obtained within a well-defined 
benchmark simulating a realistic use case. In contrast, the same modifications applied to XceptionTime 
degraded its performance, highlighting that training strategies and pre-processing can be critical performance 
factors. Overall, our findings demonstrate that UDACNN when trained using a step scheduler and Envelope(RF) 
as input, surpasses all evaluated DL techniques and even outperforms XceptionTime by $2.28$ percentage 
points while featuring $87.52\%$ fewer parameters. 

%
%
%
 \bibliographystyle{splncs04}
 \bibliography{ICPR_2026_LaTeX_Templates/references}





\end{document}